%% file: Master_Document.tex
\documentclass[authoryear,review,12pt]{elsarticle}
\usepackage[authoryear]{natbib}
\usepackage{amsmath, amssymb}

\usepackage{cases}
\usepackage[a4paper]{geometry}

\usepackage{caption}
\usepackage{subcaption}

\usepackage{eurosym}

\usepackage{booktabs}

\usepackage[nofiglist, notablist]{endfloat}

\usepackage{graphicx}
\usepackage{epstopdf}
\usepackage{rotating}
\usepackage[]{efxmpl}

\long\def\symbolfootnote[#1]#2{\begingroup%
\def\thefootnote{\fnsymbol{footnote}}\footnote[#1]{#2}\endgroup}

\makeatletter
\def\unnumfootnote{\xdef\@thefnmark{}\@footnotetext}

\begin{document}
\begin{frontmatter}

\date{}

\title{Anatomy of a Bail-In}

\long\def\symbolfootnote[#1]#2{\begingroup%
\def\thefootnote{\fnsymbol{footnote}}\footnote[#1]{#2}\endgroup}

% \author[ucd]{Thomas Conlon} %\symbolfootnote[1]{Smurfit School of Business, University College Dublin, Carysfort Avenue, Blackrock, Co. Dublin, Ireland. Email: conlon.thomas@ucd.ie}} 
% \author[ucd]{John Cotter} %\symbolfootnote[2]{Corresponding author.  Smurfit School of Business, University College Dublin, Carysfort Avenue, Blackrock, Co. Dublin, Ireland. Email: john.cotter@ucd.ie}}
% \address[ucd]{Smurfit School of Business, University College Dublin, Carysfort Avenue, Blackrock, \\ Co. Dublin, Ireland.}

\author[1]{Thomas Conlon\corref{cor1}}
\ead{conlon.thomas@ucd.ie}
\cortext[cor1]{Corresponding Author, Tel: +353-1-716 8909.}

\author[1]{John Cotter}
\ead{john.cotter@ucd.ie}

\address[1]{UCD Center for Financial Markets, \\ Smurfit Graduate School of Business, University College Dublin, Ireland}
%\cortext[cor1]{Corresponding Author}

\begin{abstract}
To mitigate potential contagion from future banking crises, the European Commission recently proposed a framework which would provide for the \textit{bail-in} of bank creditors in the event of failure. In this study, we examine this framework retrospectively in the context of failed European banks during the global financial crisis.  Empirical findings suggest that equity and subordinated bond holders would have been the main losers from the \euro 535 billion impairment losses realized by failed European banks.  Losses attributed to senior debt holders would, on aggregate, have been proportionally small, while no losses would have been imposed on depositors.  Cross-country analysis, incorporating stress-tests, reveals a divergence of outcomes with subordinated debt holders wiped out in a number of countries, while senior debt holders of Greek, Austrian and Irish banks would have required bail-in.  
\end{abstract}


\end{frontmatter}

%\unnumfootnote{Contact Information: Corresponding Author - Thomas Conlon - Email: conlon.thomas@ucd.ie Tel: +35317168909, John Cotter - Email: john.cotter@ucd.ie, Tel: +35317168900.}
\unnumfootnote{The authors would like to acknowledge the financial support of Science Foundation Ireland under Grant Number 08/SRC/FM1389. We are grateful for helpful comments and assistance from two anonymous referees, the Editor (Iftekhar Hasan), Gregory Connor, Ajay Chopra, Zhenyu Wang and participants at the FMC$^2$ Bank Resolution Conference (Dublin, Ireland, June $2013$).}

\newpage

\section{Introduction}
The credit crunch or global financial crisis, which begun in 2007, is the most severe since the great depression and has been characterised by the large number of distressed and failed systemically important financial institutions \citep{Acharya2013,Veronesi2010,Brunnermeier2009}.  The crisis has highlighted the ongoing need for a robust and consistent mechanism to allow for the resolution of failed banks \citep{Laeven2010}.  In particular, the European response to the large number of distressed banks has been fragmented and capricious \citep{Schich2012}.  Individual European Union member states took a uncoordinated approach to the crisis, re-capitalizing and nationalizing a range of domestic financial institutions \citep{Dubel2013}.  Ultimately, this approach may have contributed to financial contagion as investors had little clarity regarding the resolution mechanism to be adopted across nations, potentially resulting in a `flight-to-safety' \citep{DeBruyckere2013,Longstaff2010,Mizen2008}.

In order to mitigate potential contagion from a distressed banking sector, in $2012$ the European Commission proposed a Framework for Bank Recovery and Resolution (BRR) \citep{Commission2012}.\footnote{The BBR will be implemented from $1^{st}$ January 2015, with bail-in elements to follow from $1^{st}$ January 2016}\textsuperscript{,}\footnote{Similar frameworks have been outlined in both the United States and Great Britain, with bail-in of creditors a common feature \citep{FederalDepositInsuranceCorporation2012}.}  This was recently supplemented by the Single Resolution Mechanism (SRM), through which the ECB will apply the proposals on bank resolution, \citep{TheEuropeanCommission2013}.  Within these frameworks, failed banks will be recapitalized either through the mandatory write-down of liabilities or, alternatively, the conversion of liabilities to equity (a ‘bail-in’ of creditors or debt write-down).  It is anticipated that this mechanism will allow distressed financial institutions to continue as a going concern, while shareholders will be diluted or wiped out.  Moreover, the bail-in mechanism would help sever the link between systemically important financial institutions and the sovereign.

As outlined in the BRR, a bail-in would apply to all liabilities not backed by assets or collateral, but not to deposits protected by a deposit guarantee scheme\footnote{While one of the objectives of the resolution and recovery framework is to project depositors, deposit funding of up to \euro $100,000$ would in practise be recouped using a deposit guarantee scheme.  In effect, this results in bail-in for all failed bank depositors, with smaller depositors benefiting from a sovereign guarantee.}, short-term (inter-bank) lending or client assets.  The ordinary allocation of losses and ranking process in the event of insolvency would be followed.  Under the framework, equity holders would absorb initial losses in their entirety before any debt claim is subject to write-down.  Next, subordinated debt holders would equally share any further losses, followed by senior debt holders. Finally, depositors not protected by a deposit guarantee scheme would absorb losses.  A limited number of exempt liability holders have been proposed, including secured liabilities, trade liabilities, covered deposits, certain derivatives and short term debt with a maturity of under one month.

It is important to note that the design of a bail-in differs from contingent capital liabilities such as CoCos which provide for contingent conversion to equity in the case of bank failure \citep{Koziol2012}\footnote{Also related is the issuance of subordinated debt by banks. While subordinated debt should be priced to reflect the risk of bank failure, there are mixed views on whether subordinated debt results in market enforced discipline among financial institutions \citep{Sironi2003,Flannery1996}.}.  These securities are structured and purchased by investors on the basis of possible conversion from debt to equity, with maximum losses equivalent to the notional security face value.  A bail-in would result in mandatory conversion with the total write-down level determined by the level of banking losses.  However similarities exist, with a conversion trigger required in both cases.  In the case of instruments such as CoCos, conversion criteria have tended to be contingent upon bank capital levels deteriorating below a certain level \citep{McDonald2013,Glasserman2012}.  In contrast, a bail-in is a statutory power allowing authorities to write-down bank liabilities, with specifics regarding the trigger largely undefined at this point. 

Future funding costs of financial institutions will likely depend upon the level of anticipated write-downs that might be imposed on creditors.  However, little is known regarding the impact of a bail-in on the different liability holders.  In this paper, we retrospectively study the proportion of liabilities that authorities would have needed to bail-in to cover losses associated with the global financial crisis.  In particular, we measure the magnitude of actual impairment charges experienced by banks after 2007 and apply these to banks that required bail-out.  Finally, we perform stress test analysis to help understand the impact of considerably larger losses on creditors in the event of bail-in.

This paper contributes to the debate on the form that future resolution mechanisms will take.  The results suggest that the aggregate impairment charges of \euro 534 billion experienced by failed European banks would predominantly have impacted equity and subordinated liabilities under the proposed bail-in framework.  Cross-country analysis suggests that senior debt holders would only have incurred losses in Austria, Greece and Ireland, while depositors would not have been bailed-in within any of the states examined.  Moreover, even under stressed conditions with losses up to $20\%$ of total assets, depositors would not have required bail-in.  The findings suggest that a bail-in mechanism that largely impacts subordinated investors would help to reduce the danger of flight-to-safety, in particular limiting the impact of bank-runs by depositors.  Further, the bail-in framework should result in the returns associated with bank debt securities being linked to their explicit risk, perhaps reducing the excessive leverage associated with banks that underperformed during the crisis \citep{Beltratti2011}.  Finally, a bail-in mechanism would help to formally cut the links between the sovereign and financial institutions deemed `too big to fail', by removing the requirement for sovereign bail-out of failed financial institutions.

The paper is organized as follows: Section \ref{Outline} provides an outline of the European proposals related to resolution and bail-in of financial institutions.  Section \ref{Data} introduces the data related to failed and surviving European banks, while results from retrospectively applying the bail-in framework to European banks are described in section \ref{Results}.  The bail-in mechanism is discussed and some concluding remarks given in section  \ref{Conclusion}.

\section{European Proposals on Bank Resolution and Recovery}
\label{Outline}
In order to ensure long term financial stability and reduce the potential cost of future bank failures, the European Commission recently introduced the \textit{Framework on Bank Resolution and Recovery} \citep{Commission2012} and agreed a \textit{Single Resolution Mechanism} \citep{TheEuropeanCommission2013}.\footnote{This section provides an outline of the BRR and SRM, with particular focus on the bail-in component of each.  Considerable further detail on the resolution mechanism proposed may be found in \cite{Commission2012} and \cite{TheEuropeanCommission2013}.  Note: While the legislation underlying the SRM has yet to be ratified by the European parliament, any changes made should not greatly alter the main findings in this paper.} The resolution of a financial institution is defined as the restructuring of the institution in order to ensure the continuity of its essential functions, preserve financial stability and restore the viability of all or part of that institution \citep{Commission2012}.  Under the framework definition, a bank would become subject to resolution when
\begin{itemize}
\item
it has reached a point of distress such that there are no realistic prospects of recovery over an appropriate timeframe,
\item
all other intervention measures have been exhausted, and
\item
winding up the institution under normal insolvency proceedings would risk prolonged uncertainty or financial instability.
\end{itemize}
The framework further prescribes a range of resolution tools to be implemented dependent on the circumstances surrounding the difficulties experienced by the particular financial institution, including private sector acquisitions, `good' and `bad' banks to hold performing and toxic assets, a bridge bank to hold the assets and liabilities of failed institutions, and finally a bail-in of creditors.

In July $2013$, the European Commission subsequently proposed the single resolution mechanism (SRM) as part of the commitment to a single European banking union (the single supervisory mechanism\footnote{The single supervisory mechanism will transfer all responsibilities for the prudential supervision and authorisation of financial institutions to the European Central Bank.}).\footnote{The European parliament and council backed the SRM on $20^{th}$ March 2014, with formal ratification due to take place in April 2014.  The SRM is scheduled to enter into force on 1 January 2015, whereas bail-in and resolution functions would apply from 1 January 2016, as specified under the Bank Recovery and Resolution Directive.}    The SRM will apply the single rulebook on bank resolution proposed in the European framework on bank resolution and recovery, \citep{TheEuropeanCommission2013}.  The SRM sets out in detail the order of priority in case of write downs or conversion to equity and the assessment of the amount by which liabilities need to be converted.  The decision on when to recommend the resolution of a bank to the European Commission lies with a resolution board, made up of executive and deputy directors, representatives appointed by the Commission and ECB, together with members appointed by individual European member states.  The SRM will be able to draw on a resolution fund, funded by contributions from the banking sector and replacing national resolution funds, in the exceptional event that additional resources are required.  Banks will be required to hold minimum required eligible liabilities, determined based on an institutions size, risk and business model, to mitigate the possibility of depositor bail-in.

Details on the procedure for placing a bank into resolution under the SRM are presently being finalised.  The most recent draft regulation\footnote{Our synopsis is based on the presidency compromise text issued on November $4^{th}$ $2013$, document $15503/13$.} states that the procedure begins with an assessment by the ECB that (i) a bank is failing or likely to fail (Article 16.2a), and (ii) there is no reasonable prospect that any supervisory or private action would prevent its failure within a reasonable time-frame (Article 16.2b).  Should an institution be found to be in violation of this assessment, the finding is communicated to the European Commission and the single resolution board (Article 16.1).  Conditions under which a bank may be pronounced as failing or likely to fail are listed in Article 16.3:  where the bank is in breach of requirements for authorisation (insufficient capital); its net worth is negative; it is or will soon be unable to repay its debts; or there is a need for extraordinary public support.\footnote{While the conditions stated define the general circumstances for bank failure, they do not prescribe specific triggers to instigate a resolution process.  These might include quantitative measures such as equity capital or liquidity deteriorating below a pre-determined level.} The regulation then requires that these facts are verified by the single resolution board and reported to the European Commission.  The Commission will then make the final decision on whether to adopt the resolution recommendations.

The single resolution board will draw up a resolution scheme, one component of which may require the use of bail-in.  The resolution scheme will determine the amount by which eligible liabilities will be reduced or converted to equity.  The pecking order for bail-in adopted in the BRR is followed in the SRM regulation (Article 15):  Common Tier 1 Equity, Additional Tier 1 and Tier 2 capital, subordinated debt, unsecured debt, unsecured claims and finally uncovered deposits, and the deposit guarantee scheme in lieu of guaranteed depositors which are excluded. 

\section{Data and Bank Failures}
\label{Data}
In order to retrospectively assess the impact of bank resolution through bail-in on European banks, we first identified which banks suffered distress during the crisis and the resulting bail-out mechanism applied.  Without an explicit definition of the trigger for bail-in, the timing of an actual sovereign bail-out is a reasonable proxy for our retrospective analysis.  However, it is worth noting that as further details on the bail-in framework are worked out, a more considered bail-in trigger may emerge.

In this study, banks are divided into three groupings: Nationalized banks, which were fully nationalized by the state, re-capitalized banks, which received either preferred or ordinary share capital from government sources and surviving banks, which required no state support in the form of capital injection.\footnote{Surviving banks may have received support in other ways such as covered bond issuance, where the sovereign guaranteed bond issuance.  Given the prevalence of this mechanism in Europe throughout the crisis, these institutions have not been considered as failed for the purposes of this study.}  Information relating to the bail-out status of different financial institutions was obtained from a variety of sources, \citep{Molyneux2012,Altunbas2011,Laeven2010, Petrovic2009,Goddard2009}.  In total $15$ European banks were identified as nationalized, $66$ as re-capitalized by the sovereign and $691$ required no sovereign support.   The cross-country breakdown is given in table \ref{tab:Writedowns}.  The majority of nationalized banks were in Ireland and Great Britain, while re-capitalization occurred across a wide range of countries.

Fundamental accounting related data for European banks was obtained from Bankscope.  The data has been standardized, corresponds to IFRS accounting regulations and all data has been converted into a common Euro currency at the appropriate time synchronized exchange rate.  A range of filters are used to ensure no double counting of banks due to subsidiaries.  Where possible, the bank holding company was studied to avoid double counting.  Consolidated bank accounting data was used throughout.  Balance sheet information for individual banks was sourced from Bankscope and aggregated at country, failure status and overall European level in the following analysis.

\section{Empirical Results}
\label{Results}

To retrospectively estimate the impact of bail-in on European bank investors during the global financial crisis, we first take the distribution of failed versus surviving banks in table \ref{tab:Writedowns} and calculate their subsequent impairment charges.  Total realized impairment charges for each classification (eg. re-capitalized) were identified by summing loan write-downs, non-recurring expenses (once-off expenses inclusive of losses on credit derivatives such as CDOs) and security impairments from $2008$ to $2012$.  While further related losses may be realized in the future, this is likely the best estimate of impairment charges currently available.  Note that this measure does not contain additional losses that may be realized by special purpose vehicles set up by sovereign states to hold impaired bank assets.  However, as a robustness check against underestimation of impairments, we further stress-test the bail-in mechanism for considerably greater losses than those realized during the credit crunch.

Total realized European bank impairment charges between $2008$ and $2012$ are estimated as \euro 940 billion.  Of this total, $43.1\%$ can be attributed to the $691$ surviving banks.  The remaining $56.9\%$ or \euro 535 billion is associated with the $81$ banks that were nationalized or required re-capitalization.  Considering individual countries, the largest total impairments for failed banks were experienced in Great Britain, followed by Germany and Ireland.  In terms of banks that survived without government assistance, Spanish, British and Italian banks accounted for the majority of losses.  

\input{"Table_Writedowns.tex"}

As the bail-in mechanism would apply to all unsecured bank liabilities, it is important to gain  insight into the funding sources utilized by European banks in the lead up to the global financial crisis.  Table \ref{tab:Agg_Bal_Sheet} details the aggregate funding proportions for the European banks studied at year-end $2006$ and $2007$.\footnote{While differences in funding between $2006$ and $2007$ can be observed, little quantitative difference exists between $2006$ and previous years.  Results not shown for brevity, but are available from the authors.}  As funding choices in $2007$ may have been impacted by the global financial crisis, we focus on $2006$ funding here.  Considering all banks, customer deposits accounted for $35.2\%$ of funding, bank deposits $15.7\%$ and equity $4.5\%$.  Long term bank debt accounted for $18.9\%$ of total funding, the majority of which was senior bank debt.  Finally, other liabilities, inclusive of derivatives, non-interest liabilities, repos and trading liabilities accounted for  $25.5\%$ of liabilities in $2006$.

\input{"Table_Agg_Balance_Sheet.tex"}

Contrasting failed versus surviving banks, some differential characteristics are evident.  The level of equity funding used by nationalized banks $(2.9\%)$ and recapitalized banks $(4.2\%)$ was noticeably lower than that found for surviving banks $(5.0\%)$.  Since regulatory capital is determined relative to risk weighted assets, this suggests that failed banks reported assets with lower risk levels than surviving banks.\footnote{This observation is based on the fact that all failed European banks met the minimum regulatory risk weighted tier $1$ capital ratio in $2006$.}  Considering the mix of depository and long-term funding, banks that were subsequently nationalized are shown to have largely depended on long-term debt funding, in contrast to other banks.  In particular, the proportion of long-term debt for nationalized banks was $40.8\%$, compared to a base case of $18.9\%$.  Given this higher dependence on long-term funding, these banks were potentially more susceptible to liquidity problems. However, in the context of the bail-in mechanism proposed by the European Commission, the higher levels of debt funding imply a lower risk of depositor bail-in with losses experienced first by debt holders.

Having gained some insight into the level of bank losses associated with the crisis and the funding structure adopted by European banks, we now retrospectively examine the impact of the proposed European bail-in mechanism on bank creditors, given in table \ref{tab:Bail_In}.   To this end, we apply the BRR to the $81$ European banks deemed as having failed.  We measure the proportion of total bank liabilities that would have been bailed-in to cover realized bank losses.  In addition, we break out the allocation of bail-in losses among the different capital providers in order of seniority.   In each case, realized impairment charges from $2008-2012$ are measured as a proportion of aggregate $2006$ bank liabilities.\footnote{As noted above, balance sheets of certain banks may reflect credit crunch losses by $2007$.  For instance, Northern Rock received liquidity support from the Bank of England in September $2007$.  However, our results are qualitatively similar using $2007$ balance sheets and are available upon request.}

Considering first the aggregate bail-in impact on failed European banks, we note differing consequences for the various creditors in table \ref{tab:Bail_In}.  For recapitalized banks, only equity holders would have been bailed-in to cover losses.  In contrast nationalized banks would, on aggregate, have required bail-in of all equity and subordinated debt holders, and $6.9\%$ of senior debt holders to cover realized impairments.  These results suggest that the extreme losses sustained by banks during the global financial crisis would have predominantly resulted in losses for equity and subordinated debt holders in European banks.\footnote{\cite{Dubel2013} also finds that senior debt holders would have required bail-in in isolated cases, for a study of a small number of European banks.  Moreover, evidence is given that the level of bail-in was closely related to the delay involved in resolving an institution.}  However, the picture is not quite so straightforward when individual nations are considered.

\input{"Table_Bail_In.tex"}

Table \ref{tab:Bail_In} also retrospectively considers the impact of a bail-in on banking investors across the range of European countries considered.  While equity holders would have been the predominate losers in the majority of countries, subordinated debt holders would also have encountered losses in Austria ($100\%$ for nationalized banks), Germany ($100\%$ for both recapitalized and nationalized banks), Great Britain ($75.5\%$ for recapitalized banks), Greece ($100\%$), Ireland ($100\%$) and Portugal ($50\%$).  Senior debt holders would have encountered small losses in Germany and Austria, while large write-downs would have been required in Greece ($77.8\%$) and Ireland ($24.6\%$ for nationalized banks, $64.5\%$ for recapitalized). On aggregate, the largest losses would have been felt by investors in Irish and Greek banks, accounting for up to $23.3\%$ of total balance sheet liabilities.

The analysis so far suggests that a bail-in of creditors would not have resulted in outright decimation for bank investors.  A stress-test analysis is now performed to understand the impact of further losses, due either to underestimation of realized losses resulting from the credit crunch or the possibility of a more severe financial crisis.  Table \ref{tab:Stress} details the aggregate impact on various investors of the actual realized losses calculated for European banks, in addition to larger losses amounting to $10\%$ and $20\%$ of total assets.  In both stress-test cases, equity and subordinated debt investors would have experienced full write-down for both nationalized and recapitalized institutions.   The outcome for senior debt write-downs varies for nationalized and re-capitalized institutions.  As shown earlier, nationalized banks tended to be heavily reliant on long term debt, in particular senior debt.  This preponderance of senior debt would result in proportionally lower required write-downs.  In contrast, re-capitalized banks had greater dependence on deposits resulting in large proportional write-downs for senior debt holders.  However, even with $20\%$ losses, amounting to $500\%$\footnote{This is calculated as the level of stress test losses over the actual losses for recapitalized banks, ${20\%}/{4.0\%} = 500\%$.} of the actual realized losses from the global financial crisis, depositors would not have been bailed-in.  

\input{"Table_Stress_Test.tex"}

One major difficulty in developing an efficient bail-in mechanism is the danger of contagion from bail-in of a single institution, due to other financial institutions holding outstanding debt of the failed institution.\footnote{We are grateful to an anonymous referee for pointing out this obstacle.} \cite{Zhou2012} suggest that an incorrectly structured bail-in mechanism may have the impact of shifting risk to other parts of the financial sector.  $17\%$ of Euro area bank debt was purchased by other financial institutions in $2010$, while insurance companies hold between $20\%$ and $30\%$ of their investment portfolios in bank debt\footnote{While the impact of contagion from a bail-in on insurance companies would be significant, measuring the magnitude of this impact is outside the scope of this paper.}, \citep{Zhou2012}.  Our stress-test findings for nationalized and recapitalized European banks suggest that in the case of $20\%$ losses, consistent with the proportion of bank debt held by other banks, depositors would not have required bail-in on aggregate.  As these losses would not just be limited to failed or recapitalized banks, we further stress-test surviving banks to a level of $20\%$ losses.  In this extreme scenario, with all banks requiring writedowns to a level of $20\%$ of total assets,  equity and subordinated debt holders would have been fully written down, while $77\%$ of senior debt would have been bailed-in.   Even in this case of widespread losses across European banks, perhaps a consequence of contagion,  depositors would not on aggregate have required bail-in across failed or surviving banks.

These stress test results are suggestive of a large potential impact on bank funding costs, given the reality of heavy losses for debt and equity investors in the face of severe future banking crises.  Moreover, they support the idea of the larger capital base in the Basel III proposals and the introduction of hybrid and contingent capital as an additional buffer against asset risks.

\section{Conclusions and Discussion}
\label{Conclusion}
The vast repercussions for sovereign balance sheets from the global financial crisis has led authorities to seek ways to remove the implicit link between the financial system and the sovereign.  As part of a wider framework on bank resolution, the European Commission have proposed bail-in of banking investors in the context of future banking crises.  By imposing losses on the creditors of financial institutions, the impact on public finances from banking distress should be substantially reduced.  Moreover, the explicit statement of intent regarding creditor write-downs should have a positive impact on the debt issuance of sovereign states in the short term, important in the context of peripheral European states.

The findings outlined in this paper suggest large realized European banking impairments associated with the global financial crisis. Existence of a bail-in mechanism would have predominantly impacted equity and subordinated debt investors in European banks.  Senior debtors would have experienced write-downs in a limited number of nations, including Greece, Austria, Germany and Ireland. Stress-test analysis demonstrates that impairments of up to $20\%$ of total assets would have resulted in losses of up to $96\%$ for senior debt investors.  Even in this extreme scenario depositors would not have experienced write-downs, limiting the danger of a `flight-to-safety' and associated contagion due to bail-in. 

The results detailed, in particular those outlining stress-test analysis, suggest that aggregate bank borrowing costs for financial institutions would likely rise in the face of potentially large bail-in write-downs.  In keeping with previous findings linking bank risk and subordinated debt yields, \citep{Sironi2003,Flannery1996}, investors are likely to weigh up the likelihood of an investment in a financial institution being subject to mandatory write-down and expect a return commensurate with these risks. Given the link between bank performance and leverage throughout the global financial crisis, \citep{Beltratti2011}, this would have the effect of forcing banks to have less leveraged balance sheets, more appropriate to the riskiness of their asset portfolios.\footnote{Bank management have noted that increased levels of equity capital would have an adverse effect on the ability of banks to lend, a theory that has been strongly criticized, \citep{Admati2010}.  Further, under certain circumstances high leverage may actually have a positive effect on banks via socially useful liquidity creation to financially constrained firms and households \citep{DeAngelo2013}.}   

%This contrasts to the current regime where investors are willing to lend to financial institutions in expectation of returns normally associated with very low risk investments (sovereign debt for instance), under the assumption that systemically important financial institutions will be bailed-out in times of distress.  
%
%The vast differences in bank performance throughout the credit crunch has been linked to the degree of leverage applied \citep{Beltratti2011}.   Equity and subordinated debt investors have been widely shown to discern between banks with varying levels of risk \citep{Sironi2003,Flannery1996}.  Similarly, following the introduction of a bail-in framework senior debt investors also would link the cost of borrowing to the actual risk involved. 

The current draft SRM regulation requires banks to hold a minimum level of `bail-inable' assets.   This may necessitate an alteration in individual bank capital structures in order to meet these requirements.  \cite{Zhou2012} raises the concern that banks may decide to shift their borrowings towards short-term and secured financing under the bail-in proposal.  A shift in bank borrowing towards short-term, bail-in exempt securities could be counter-productive, actually adding to the dangers of financial contagion.   However, the proposal to ensure banks hold a minimum quantity of `bail-inable' securities should mitigate this concern.

A variety of potential shortfalls surrounding the proposed European bail-in mechanism have been highlighted.  For a large systemically important financial institution which is globally active, domestic European bail-in laws would not necessarily be recognised in other jurisdictions, which might result in ambiguity as to which assets could be bailed-in, \citep{Gleeson2012}.  In addition, countries which are early adopters of a bail-in mechanism might disadvantage their financial institutions in terms of higher costs of funding, \citep{Schich2012}.  Moreover, the resolution process as currently proposed is a cumbersome process requiring approval from a variety of parties, potentially slowing the process and creating uncertainty surrounding an institution.  The exact treatment under bail-in of other liabilities such as repos, derivatives and trading liabilities (which account for a large proportion of balance sheet liabilities, table \ref{tab:Agg_Bal_Sheet}) is currently unclear, adding potential ambiguity to the bail-in process.

Another major concern with the proposed resolution process and the associated bail-in mechanism is the ambiguity regarding the trigger resulting in resolution.\footnote{The ECB, in its role in the single supervisory mechanism, seems to be cognizant of this poential issue.  Mario Draghi, President of the ECB, stated during his opening speech at the European Banking Congress ``The key to an effective resolution regime is that it creates legal certainty, consistency and predictability, thus helping to avoid ad-hoc solutions''.}   While the SRM details general conditions under which banks would be deemed to have failed, an explicit indicator regarding what constitutes a bank failure would benefit both regulators and bank investors.  Indeed without explicit quantitative clarity on the trigger for creditor write-downs, investors may require a risk premium in compensation.  Linking the trigger to the leverage ratio or tier one capital is one method that has been applied for contingent capital \citep{Glasserman2012}.  However, a range of market based measures linking the trigger to equity levels in addition to yields on long-term debt may be appropriate in the case of bail-in \citep{Sundaresan2013}.  This would allow debt investors to explicitly determine the risk of bail-in from market related borrow costs.  It is important to note that the establishment of a bail-in framework for failed banks does not preclude the introduction of contingent convertible instruments as a means of bank funding.  In fact, our analysis suggests that the introduction of an extra layer of hybrid capital would act as a buffer for senior claims, altering the impact of a bail-in framework on higher ranked securities.  Finally, a pre-insolvency resolution trigger would allow for a prompt and effective reaction to bank distress, potentially reducing the overall cost to investors and society, \citep{Zhou2012}.  There are, however, also dangers associated with a pre-insolvency trigger, with potential for unnecessary bail-in.  

Introduction of a European bail-in mechanism may alter the intrinsic funding costs faced by European banks.  A survey by investment bank J.P. Morgan suggested that the introduction of bail-in would result in an expected increase of $87$ basis points in the long-term debt yield for a single A rated bank, with an estimated $3-4$ notch downgrade on Moody's rating scale.\footnote{J.P. Morgan survey of 55 European Banks, \textit{The Great Bank Downgrade} January 2011.}  The European Commission has estimated an overall increase in bank funding costs including short-term debt of $31.6$ basis points, \citep{Commission'2012a}.  The Commission has further calculated that the combined cost of increasing capital requirements to meet the $10.5\%$ Basel III minimum and simultaneously introducing bail-in would range between $0.5\%$ and $1.2\%$ of GDP per annum.   In contrast, European GDP dropped by $4.4\%$ during $2009$, while experiencing total growth of $1.5\%$ between $2007$ and $2012$.\footnote{Source:  Eurostat, Euro Area real GDP growth rate (17 countries).}  This suggests the balance between the costs of financial stability and the potential cost of prudence should be carefully considered.

\newpage

\vspace{10mm}
\begin{center}
\bibliographystyle{authordate1new}
%\bibliography{D:/Dropbox/Research/Bibtex/library}
\bibliography{references}

\end{center}

%  The following lines expand the size of the top and bottom margins for the tables and figures only.
 \addtolength{\textheight}{2.5cm}
\addtolength{\voffset}{-1.5cm}

\end{document}

%% file: Table_Writedowns.tex
% Table generated by Excel2LaTeX from sheet 'Sheet3'
\begin{table}[htbp]
\begin{changemargin}{-1cm}{-1cm}
\footnotesize
  \centering
\renewcommand{\arraystretch}{1.2}
    \begin{tabular}{l | ccc | rrrr}
    \hline
          & \multicolumn{3}{c|}{\textbf{Number of Banks 2007}} & \textbf{} & \multicolumn{3}{c}{\textbf{Impairment Charges 2008-2012 (\euro m)}} \\
    \textbf{Country} & \textbf{Capital} & \textbf{Nationalized} & \textbf{Surviving} & \textbf{ } & \textbf{Capital } & \textbf{Nationalized } & \textbf{Surviving } \\
    \hline
    \textbf{Austria} & 2     & 2     & 20    &       & 13,952 & 3,959 & 2,679 \\
    \textbf{Belgium} & 3     & 0     & 6     &       & 32,600 & 0     & 205 \\
    \textbf{Germany} & 6     & 1     & 516   &       & 89,115 & 8,359 & 27,734 \\
    \textbf{Denmark} & 1     & 2     & 19    &       & 10,804 & 0     & 3,845 \\
    \textbf{Spain} & 0     & 1     & 28    &       & 0     & 0     & 141,125 \\
    \textbf{Finland} & 0     & 0     & 4     &       & 0     & 0     & 87 \\
    \textbf{France*} & 35    & 0     & 13    &       & 0     & 56,800 & 8,340 \\
    \textbf{Great Britain} & 3     & 4     & 27    &       & 130,141 & 4,313 & 117,876 \\
    \textbf{Greece} & 7     & 0     & 2     &       & 33,967 & 0     & 715 \\
    \textbf{Ireland} & 2     & 4     & 0     &       & 45,844 & 42,247 & 0 \\
    \textbf{Italy} & 2     & 0     & 40    &       & 34,432 & 0     & 73,757 \\
    \textbf{Netherlands} & 4     & 1     & 1     &       & 22,997 & 0     & 11,505 \\
    \textbf{Portugal} & 1     & 0     & 6     &       & 0     & 5,245 & 7,049 \\
    \textbf{Sweden} & 0     & 0     & 9     &       & 0     & 0     & 10,719 \\
    \textbf{} &       &       &       &       &       &       &  \\
    \textbf{Total} & 66    & 15    & 691   &       & 413,852 & 120,923 & 405,636 \\
    \textbf{} & \textbf{} & \textbf{} & \textbf{} & \textbf{} & \textbf{} & \textbf{} & \textbf{} \\
    \textbf{Grand Total} & \textbf{} & \textbf{} & \textbf{772} & \textbf{} & \textbf{} & \textbf{} & \textbf{940,411} \\
    \bottomrule
    \end{tabular}
    \caption{\footnotesize \textbf{Breakdown of bank status and total impairment charges (2008-2012)} \\ European bank failures between 2008 and 2009 are shown, broken out by country according to `Capital' - government capital injection required, `Nationalized' - bank nationalized and `Surviving' - survived without government assistance.  Total realized impairment charges are calculated for each bank as the sum of loan writedowns and non-recurring expenses between 2008 and 2012.  All data is sourced from Bankscope and given in millions of Euro. *Note: The French bank Credit Agricole is treated as a collection of separate cooperative institutions in this study due to missing aggregate data for the combined entity.}
      \label{tab:Writedowns}
    \end{changemargin}
\end{table}

%% file: Table_Agg_Balance_Sheet.tex
% Table generated by Excel2LaTeX from sheet 'Sheet2'
\begin{table}[htbp]
  \begin{changemargin}{-1cm}{-1cm}
\footnotesize
  \centering

    \begin{tabular}{l | cccc | cccc}
    \hline
          & \multicolumn{4}{c |}{\textbf{2006}}      & \multicolumn{4}{c|}{\textbf{2007}}       \\
    \hline
    \textbf{Liability Type} & \textbf{Capital} & \textbf{Nat.} & \textbf{Sur.} & \textbf{All} & \textbf{Capital } & \textbf{Nat.} & \textbf{Sur.}  & \textbf{All} \\
    \textbf{Total Customer Deposits} &  32.1\% & 27.7\% & 38.6\% & 35.2\% & 31.2\% & 20.5\% & 36.7\% & 33.5\%  \\
    \textbf{Deposits from Banks} &  19.4\% & 14.9\% & 12.3\% & 15.7\% & 17.8\% & 21.5\% & 10.6\% & 14.4\%   \\
   \textbf{Total Long Term Debt} & 16.3\% & 40.8\% & 19.9\% & 18.9\% & 15.1\% & 36.3\% & 17.8\% & 17.3\%   \\
   \textit{Senior Debt} &  13.9\% & 39.3\% & 16.0\% & 15.8\% & 12.3\% & 34.5\% & 15.0\% & 14.5\%   \\
    \textit{Subordinated Debt} &  1.6\% & 1.5\% & 1.6\% & 1.6\% & 1.6\% & 1.7\% & 1.5\% & 1.6\%   \\
    \textit{Other Funding} &  1.9\% & 0.0\% & 3.4\% & 2.6\% & 1.1\% & 0.1\% & 1.3\% & 1.1\%   \\
    \textbf{Other Liabilities}  &  27.9\% & 13.1\% & 24.2\% & 25.5\% & 31.4\% & 16.8\% & 29.4\% & 29.8\% \\
   \textbf{} &       &       &       &       &       &       &       &         \\
    \textbf{Total Liabilities} & \textbf{95.8\%} & \textbf{97.1\%} & \textbf{95.0\%} & \textbf{95.5\%} & \textbf{96.0\%} & \textbf{96.2\%} & \textbf{95.1\%} & \textbf{95.5\%}   \\
    \textbf{Total Equity} &  \textbf{4.2\%} & \textbf{2.9\%} & \textbf{5.0\%} & \textbf{4.5\%} & \textbf{4.0\%} & \textbf{3.8\%} & \textbf{4.9\%} & \textbf{4.5\%}  \\
    \hline
    \end{tabular}

    \caption{\footnotesize \textbf{Aggregate Balance Sheet Liabilities for European Union Banks 2006-2007} \\ 
    An aggregate European Union bank balance sheet is determined by summing over all liabilities.  Proportions are then found as a percentage of total balance sheeet liabilities for each of 2006 and 2007.  Banks are categorized as `Capital' - government re-capitalization required, `Nat.' - nationalized, `Sur.' - Survived without government assistance and `All' - All banks.  Other liabilities include derivatives, non-interest, repos and trading liabilities. }
      \label{tab:Agg_Bal_Sheet}
      \end{changemargin}
\end{table}

%% file: Table_Bail_In.tex
\begin{table}[htbp]
  \begin{changemargin}{-1cm}{-1cm}
  \centering
  \footnotesize
\renewcommand{\arraystretch}{1.2}

    \begin{tabular}{llc | cccc}
    \hline
%    \textbf{} & \textbf{} & \textbf{} & \multicolumn{4}{c}{\textbf{Impairment Charge Writedowns (2008-2012) as a Proportion of}} \\

    \textbf{} & \textbf{} & \multicolumn{1}{c}{\textbf{}} & \textbf{Total} & \textbf{Subordinated} & \textbf{Senior} & \textbf{Total} \\
        \textbf{Country} & \textbf{Status} & \multicolumn{1}{c}{\textbf{Number}} & \textbf{Equity} & \textbf{Debt} & \textbf{Debt} & \textbf{Liabilities \& Equity} \\
     \hline
    \textbf{Austria} & \textbf{Cap.} & \multicolumn{1}{c}{2} & 100.0\% & 5.7\% & 0.0\% & 5.6\% \\
    \textbf{} & \textbf{Nat.} & \multicolumn{1}{c}{2} & 100.0\% & 100.0\% & 11.2\% & 12.8\% \\
    \multicolumn{7}{c}{}\\
    \textbf{Belgium} & \textbf{Cap.} & \multicolumn{1}{c}{3} & 63.2\% & 0.0\% & 0.0\% & 2.1\% \\
        \multicolumn{7}{c}{}\\
    \textbf{Germany} & \textbf{Cap.} & \multicolumn{1}{c}{6} & 100.0\% & 100.0\% & 2.7\% & 5.1\% \\
    \textbf{} & \textbf{Nat.} & \multicolumn{1}{c}{1} & 100.0\% & 100.0\% & 3.2\% & 5.2\% \\
        \multicolumn{7}{c}{}\\
    \textbf{Denmark} & \textbf{Cap.} & \multicolumn{1}{c}{1} & 84.6\% & 0.0\% & 0.0\% & 3.0\% \\
    \textbf{} & \textbf{Nat.*} & \multicolumn{1}{c}{2} & 0.0\% & 0.0\% & 0.0\% & 0.0\% \\
        \multicolumn{7}{c}{}\\
    \textbf{Spain} & \textbf{Nat.*} & \multicolumn{1}{c}{1} & 0.0\% & 0.0\% & 0.0\% & 0.0\% \\
        \multicolumn{7}{c}{}\\
    \textbf{France} & \textbf{Cap.**} & \multicolumn{1}{c}{35} & 39.0\% & 0.0\% & 0.0\% & 1.9\% \\
        \multicolumn{7}{c}{}\\
    \textbf{Great Britain} & \textbf{Cap.} & \multicolumn{1}{c}{3} & 100.0\% & 75.5\% & 0.0\% & 5.6\% \\
    \textbf{} & \textbf{Nat.} & \multicolumn{1}{c}{4} & 85.4\% & 0.0\% & 0.0\% & 2.0\% \\
        \multicolumn{7}{c}{}\\
    \textbf{Greece} & \textbf{Cap.} & \multicolumn{1}{c}{7} & 100.0\% & 100.0\% & 77.8\% & 20.8\% \\
        \multicolumn{7}{c}{}\\
    \textbf{Ireland} & \textbf{Cap.} & \multicolumn{1}{c}{2} & 100.0\% & 100.0\% & 24.6\% & 13.4\% \\
    \textbf{} & \textbf{Nat.} & \multicolumn{1}{c}{4} & 100.0\% & 100.0\% & 64.5\% & 23.3\% \\
        \multicolumn{7}{c}{}\\
    \textbf{Italy} & \textbf{Cap.} & \multicolumn{1}{c}{2} & 57.0\% & 0.0\% & 0.0\% & 5.6\% \\
        \multicolumn{7}{c}{}\\
    \textbf{Netherlands} & \textbf{Cap.} & \multicolumn{1}{c}{4} & 52.6\% & 0.0\% & 0.0\% & 1.8\% \\
    \textbf{} & \textbf{Nat.*} & \multicolumn{1}{c}{1} & 0.0\% & 0.0\% & 0.0\% & 0.0\% \\
        \multicolumn{7}{c}{}\\
    \textbf{Portugal} & \textbf{Cap.} & \multicolumn{1}{c}{1} & 100.0\% & 49.9\% & 0.0\% & 5.5\% \\
    \textbf{} & \textbf{} & \multicolumn{1}{c}{} &       &       &       &  \\
        \multicolumn{7}{c}{}\\
        \hline
    \textbf{Europe All} & \textbf{Nat.} & \multicolumn{1}{c}{15} & 100.0\% & 100.0\% & 6.9\% & 7.1\% \\
    \textbf{} & \textbf{Cap.} & \multicolumn{1}{c}{66} & 97.1\% & 0.0\% & 0.0\% & 4.0\% \\
    \hline
    \end{tabular}
      \caption{\footnotesize \textbf{Proportion of liabilities required for bail-in to cover writedown losses by country.} \\ This table measures the level of `bail-in' that would have been required by EU banks in order to cover losses from the global financial crisis.  Total realized impairment charges are calculated for each bank as the sum of loan writedowns and non-recurring expenses between 2008 and 2012.  The proportion of each balance sheet liability required to be written down to cover these losses is then calculated. * In both Spain and Denmark nationalized banks were incorporated into a `bad-bank' resulting in no impairment charges detailed in accounts. ** The French bank Credit Agricole is treated as a collection of separate cooperative institutions in this study due to missing aggregate data for the combined entity.}
  \label{tab:Bail_In}
  \end{changemargin}
\end{table}

%% file: Table_Stress_Test.tex
% Table generated by Excel2LaTeX from sheet 'Sheet2'
\begin{table}[htbp]
\begin{changemargin}{-1cm}{-1cm}
\footnotesize
  \centering

    \begin{tabular}{lcccc}
    \toprule
    \textbf{Status} & \textbf{Writedown} & \textbf{Total Equity} & \textbf{Subordinated Debt} & \textbf{Senior Debt} \\
    \midrule
    \multicolumn{5}{c}{\textit{Actual Realized Impairments}}  \\
    \textbf{Nationalized} & 7.1\% & 100\% & 100\% & 6.9\% \\
    \textbf{Capital} & 4.0\% & 97\%  & 0\%   & 0.0\% \\
    \textbf{Surviving} & 3.37\% & 64\% & 0\% & 0.0\% \\  
        \multicolumn{5}{c}{\textit{Stress Test - Losses = 10\% of Aggregate Bank Assets}}  \\
    \textbf{Nationalized} & 10\%  & 100\% & 100\% & 13.4\% \\
    \textbf{Capital} & 10\%  & 100\% & 100\% & 27.5\% \\
    \textbf{Surviving} & 10\% & 100\% & 100\% & 18\% \\
        \multicolumn{5}{c}{\textit{Stress Test - Losses = 20\% of Aggregate Bank Assets} } \\
    \textbf{Nationalized} & 20\%  & 100\% & 100\% & 37.9\% \\
    \textbf{Capital} & 20\%  & 100\% & 100\% & 96.3\% \\
    \textbf{Surviving} & 20\% & 100\% & 100\% & 77\% \\
    \bottomrule
    \end{tabular}

    \caption{\footnotesize \textbf{Stress test analysis: Bail-in losses under extreme adverse conditions.} \\ This table stress tests the level of `bail-in' that would have been required by EU banks in order to cover losses from the global financial crisis.  Total realized impairment charges are calculated for each bank as the sum of loan write-downs and non-recurring expenses between 2008 and 2012. In addition, losses of $10\%$ and $20\%$ of aggregate assets are examined. The proportion of each balance sheet liability required to be written down to cover these losses is then calculated.}
        \label{tab:Stress}
    \end{changemargin}

\end{table}

%% file: Master_Document.bbl
\begin{thebibliography}{}

\bibitem[\protect\citename{Acharya, }2013]{Acharya2013}
Acharya, V.V. (2013).
\newblock {Understanding financial crises: Theory and evidence from the crisis
  of 2007-2008}.
\newblock {\em NYU Stern School of Business Working Paper}.

\bibitem[\protect\citename{Admati {\em et~al.}, }2010]{Admati2010}
Admati, A., DeMarzo, P., Hellwig, M., \ Pfleiderer, P. (2010).
\newblock {Fallacies, irrelevant facts, and myths in the discussion of capital
  regulation: Why bank equity is not expensive}.
\newblock {\em Stanford Graduate School of Business Working Paper}.

\bibitem[\protect\citename{Altunbas {\em et~al.}, }2011]{Altunbas2011}
Altunbas, Y., Manganelli, S., \ Marques-Ibanez, D. (2011).
\newblock {Bank risk during the financial crisis: do business models matter?}
\newblock {\em European Central Bank Working Paper}, {\bf No. 1394}.

\bibitem[\protect\citename{Beltratti and Stulz, }2012]{Beltratti2011}
Beltratti, A., \ Stulz, R.~M. (2012).
\newblock {The credit crisis around the globe: Why did some banks perform
  better?}
\newblock {\em Journal of Financial Economics}, {\bf 105}(1), 1--17.

\bibitem[\protect\citename{Brunnermeier, }2009]{Brunnermeier2009}
Brunnermeier, M.K. (2009).
\newblock {Deciphering the liquidity and credit crunch 2007-2008}.
\newblock {\em Journal of Economic Perspectives}, {\bf 23}(1), 77--100.

\bibitem[\protect\citename{{De Bruyckere} {\em et~al.}, }2013]{DeBruyckere2013}
{De Bruyckere}, V., Gerhardt, M., Schepens, G., \ {Vander Vennet}, R. (2013).
\newblock {Bank/sovereign risk spillovers in the European debt crisis}.
\newblock {\em Journal of Banking \& Finance}, {\bf 37}(12), 4793--4809.

\bibitem[\protect\citename{DeAngelo and Stulz, }2013]{DeAngelo2013}
DeAngelo, H., \ Stulz, R. (2013).
\newblock {Why high leverage is optimal for banks}.
\newblock {\em Fisher College of Business Working Paper}.

\bibitem[\protect\citename{D\"{u}bel, }2013]{Dubel2013}
D\"{u}bel, H-J. (2013).
\newblock {The capital structure of banks and practice of bank restructuring}.
\newblock {\em Study commissioned by Center for Financial Studies, University
  of Frankfurt.}

\bibitem[\protect\citename{{Federal Deposit Insurance Corporation} and {Bank of
  England}, }2012]{FederalDepositInsuranceCorporation2012}
{Federal Deposit Insurance Corporation}, \ {Bank of England}. (2012).
\newblock {Resolving globally active, systemically important, financial
  institutions}.
\newblock {\em Working Paper}.

\bibitem[\protect\citename{Flannery and Sorescu, }1996]{Flannery1996}
Flannery, M.J., \ Sorescu, S.M. (1996).
\newblock {Evidence of bank market discipline in subordinated debenture yields:
  1983-1991}.
\newblock {\em Journal of Finance}, {\bf 51}(4), 1347--1377.

\bibitem[\protect\citename{Glasserman and Nouri, }2012]{Glasserman2012}
Glasserman, P., \ Nouri, B. (2012).
\newblock {Contingent capital with a capital-ratio trigger}.
\newblock {\em Management Science}, {\bf 58}(10), 1816--1833.

\bibitem[\protect\citename{Gleeson, }2012]{Gleeson2012}
Gleeson, S. (2012).
\newblock {Legal aspects of bank bail-ins}.
\newblock {\em London School of Economics Financial Markets Group Working Paper
  Series}.

\bibitem[\protect\citename{Goddard {\em et~al.}, }2009]{Goddard2009}
Goddard, J., Molyneux, P., \ Wilson, J.O.S. (2009).
\newblock {The financial crisis in Europe: Evolution, policy responses and
  lessons for the future}.
\newblock {\em Journal of Financial Regulation and Compliance}, {\bf 17}(4),
  362--380.

\bibitem[\protect\citename{Koziol and Lawrenz, }2012]{Koziol2012}
Koziol, C., \ Lawrenz, J. (2012).
\newblock {Contingent convertibles. Solving or seeding the next banking
  crisis?}
\newblock {\em Journal of Banking \& Finance}, {\bf 36}(1), 90--104.

\bibitem[\protect\citename{Laeven and Valencia, }2010]{Laeven2010}
Laeven, L., \ Valencia, F. (2010).
\newblock {Resolution of banking crises: The good, the bad, and the ugly}.
\newblock {\em IMF Working Paper}.

\bibitem[\protect\citename{Longstaff, }2010]{Longstaff2010}
Longstaff, F.A. (2010).
\newblock {The subprime credit crisis and contagion in financial markets}.
\newblock {\em Journal of Financial Economics}, {\bf 97}(3), 436--450.

\bibitem[\protect\citename{McDonald, }2013]{McDonald2013}
McDonald, Robert~L. (2013).
\newblock {Contingent capital with a dual price trigger}.
\newblock {\em Journal of Financial Stability}, {\bf 9}(2), 230--241.

\bibitem[\protect\citename{Mizen, }2008]{Mizen2008}
Mizen, P. (2008).
\newblock {The credit crunch of 2007-2008 : A discussion of the background,
  market reactions, and policy responses}.
\newblock {\em Federal Reserve Bank of St. Louis Review}, {\bf Sept.-Oct.},
  531--568.

\bibitem[\protect\citename{Molyneux {\em et~al.}, }2011]{Molyneux2012}
Molyneux, P., Schaeck, K., \ Zhou, T.M. (2011).
\newblock {‘Too systemically important to fail ’ in banking}.
\newblock {\em Bangor Business School Working Paper}.

\bibitem[\protect\citename{Petrovic and Tutsch, }2009]{Petrovic2009}
Petrovic, A., \ Tutsch, R. (2009).
\newblock {National rescue measures in response to the current financial
  crisis}.
\newblock {\em ECB Legal Working Paper Series}, {\bf 8}.

\bibitem[\protect\citename{Schich and Kim, }2012]{Schich2012}
Schich, S., \ Kim, B-H. (2012).
\newblock {Developments in the value of implicit guarantees for bank debt: The
  role of resolution regimes and practices}.
\newblock {\em OECD Journal: Financial Market Trends}, {\bf 2012}(2), 1--31.

\bibitem[\protect\citename{Sironi, }2003]{Sironi2003}
Sironi, A. (2003).
\newblock {Testing for market discipline in the European banking industry:
  Evidence from subordinated debt issues}.
\newblock {\em Journal of Money, Credit and Banking}, {\bf 35}(3), 443--472.

\bibitem[\protect\citename{Sundaresan and Wang, }2013]{Sundaresan2013}
Sundaresan, S., \ Wang, Z. (2013).
\newblock {On the design of contingent capital with market trigger}.
\newblock {\em Journal of Finance}, {\bf To Appear}.

\bibitem[\protect\citename{{The European Commission}, }2012a]{Commission2012}
{The European Commission}. (2012a).
\newblock {Directive of the European parliament and of the council establishing
  a framework for the recovery and resolution of credit institutions and
  investment firms.}
\newblock {\em June 2012}.

\bibitem[\protect\citename{{The European Commission}, }2012b]{Commission'2012a}
{The European Commission}. (2012b).
\newblock {Impact assessment accompanying the framework for the recovery and
  resolution of credit institutions and investment firms}.
\newblock {\em June 2012}.

\bibitem[\protect\citename{{The European Commission},
  }2013]{TheEuropeanCommission2013}
{The European Commission}. (2013).
\newblock {Proposal for a regulation of the European Parliament and of the
  council establishing uniform rules and a uniform procedure for the resolution
  of credit institutions and certain investment firms in the framework of a
  single resolution mechanism.}
\newblock {\em July 2013}.

\bibitem[\protect\citename{Veronesi and Zingales, }2010]{Veronesi2010}
Veronesi, P., \ Zingales, L. (2010).
\newblock {Paulson's gift}.
\newblock {\em Journal of Financial Economics}, {\bf 97}(3), 339--368.

\bibitem[\protect\citename{Zhou {\em et~al.}, }2012]{Zhou2012}
Zhou, J., Rutledge, V., Bossu, W., Dobler, M., Jassaud, N., \ Moore, M. (2012).
\newblock {From Bail-out to Bail-in: Mandatory Debt Restructuring of Systemic
  Financial Institutions}.
\newblock {\em IMF Staff Dicussion Note}, {\bf SDN/12/03}.

\end{thebibliography}
